\documentclass{article}

\usepackage{graphicx}

\usepackage{multirow}
\usepackage{hyperref}

\usepackage[sorting=none]{biblatex}
\usepackage{amsmath}
\usepackage{amssymb}
\usepackage{dsfont}

\usepackage{authblk}

\newcommand{\bi}[1]{\boldsymbol{#1}}
\newcommand{\bomega}{\boldsymbol{\omega}}
\newcommand{\bGamma}{\boldsymbol{\Gamma}}
\newcommand{\epmatrix}[1]{\begin{pmatrix}#1\end{pmatrix}}
\newcommand{\ematrix}[1]{\begin{matrix}#1\end{matrix}}
\newcommand{\eqalign}[1]{\begin{aligned}#1\end{aligned}}
\let\amscases\cases
\makeatletter
\def\cases{\@ifnextchar\bgroup\plaincases\amscases}
\def\plaincases#1{\begin{cases}#1\end{cases}}
\makeatother

\begin{document}

\title{Direct coupling of nonlinear integrated cavities for all-optical reservoir computing}

\author{Ivan Boikov$^{1, \footnote{Electronic mail: ivan.boikov@thalesgroup.com}}$, Daniel Brunner$^{2}$ and Alfredo De Rossi$^{1}$}

\affil{$^{1}$Thales Research \& Technology, Palaiseau Cedex, 91767, France}
\affil{$^{2}$FEMTO-ST Institute / Optics Department, CNRS \& University Bourgogne Franche-Comté, Besançon Cedex, 25030, France}

\date{20 July 2023}

\maketitle

\begin{abstract}
	We consider theoretically a network of directly coupled optical microcavities to implement a space-multiplexed optical neural network in an integrated nanophotonic circuit.
	Nonlinear photonic network integrations based on direct coupling ensures a highly dense integration, reducing the chip footprint by several orders of magnitude compared to other implementations.
	Different nonlinear effects inherent to such microcavities are studied when used for realizing an all-optical autonomous computing substrate, here based on the reservoir computing concept.
	We provide an in-depth analysis of the impact of basic microcavity parameters on computational metrics of the system, namely, the dimensionality and the consistency.
	Importantly, we find that differences between frequencies and bandwidths of supermodes formed by the direct coupling is the determining factor of the reservoir's dimensionality and its scalability.
	The network's dimensionality can be improved with frequency-shifting nonlinear effects such as the Kerr effect, while two-photon absorption has an opposite effect.
	Finally, we demonstrate in simulation that the proposed reservoir is capable of solving the Mackey-Glass prediction and the optical signal recovery tasks at GHz timescale.
\end{abstract}


\section{Introduction}

Artificial Neural Networks (ANNs) are a computing approach different from the conventional algorithmic concepts.
They are inspired by the most basic principles of human brains, where numerous neurons are connected with synapses to create a network.
Currently, ANNs are mostly implemented on the von Neumann architecture, i.e. on digital computers, and, while successful in many fields, this approach has its own drawbacks, mainly a data bottleneck inherent to the architecture as well as significant power requirements.
In addition, the latency between an input and a computed output presents a major concern in some applications.
One example is optical communications, where an exponentially growing demand for internet connectivity requires increasingly faster processing of optical signals~\cite{argyris2022}.
A conventional approach is based on the processing with digital algorithms~\cite{ip2008, guiomar2012, mecozzi2016} or more recent ANN-inspired solutions~\cite{hager2018}.
However, at higher bandwidths, an efficient processing becomes complicated.
The development of ANNs in a physical layer can help overcome this challenge~\cite{shastri2017, brunner2019}.
Besides analog electronics~\cite{csaba2020}, the manipulation of optical signals is highly relevant when a large bandwidth operation is important, as was demonstrated in the field of microwave photonics~\cite{marpaung2019}.
For implementing such computing systems, integrated photonics is a promising platform that provides compactness and intrinsic nonlinear effects necessary for efficient and scalable ANN in hardware~\cite{shen2017, feldmann2021}.

Reservoir Computing (RC) has attracted a lot of interest in the in-hardware computing field as it provides an implementation-friendly ANN topology with competitive performance and a simplistic training procedure~\cite{verstraeten2007}.
Reservoir Computing on integrated photonic platforms has been demonstrated in simulations~\cite{mesaritakis2013, denis2018} and experimentally~\cite{vandoorne2014, sackesyn2021}.
These works considered optical microcavities or splitters as neurons with long intra-neural connections, i.e. with a non-negligible coupling delay.
This was required to match the system's internal transient timescales to available electronics~\cite{vandoorne2014}.
However, these delay lines are costly in terms of a chip footprint and optical losses and limit scalability.
As an example, a 280~ps delay requires 2~cm of silicon waveguide~\cite{vandoorne2014}, which results in approximately one neuron per square millimeter of the chip surface.

In this study, we consider evanescent, i.e. direct, coupling between adjacent cavities as an alternative to create maximally compact integrated all-optical computing systems.
We propose a RC consisting of a rectangular grid of directly coupled nonlinear microcavities and study the impact of cavity parameters and reservoir geometry on its basic computational properties, namely the dimensionality~\cite{dambre2012, skalli2022} and the consistency~\cite{uchida2004}.
We found that a longer photon lifetime allows for a better scalability due to more variability between coupling-induced supermodes of the system.
%
Importantly, we found that the nonlinear effects can impact computing performance in opposite directions, where two-photon absorption (TPA) reduces the reservoir's dimensionality, while free carrier dispersion (FCD) and the optical Kerr effect increase the dimensionality.
Moreover, unlike TPA, a sufficiently strong FCD and Kerr effect do induce chaotic dynamics in this purely dissipative network without gain.
Finally, we evaluate the RC's performance in Mackey-Glass prediction and optical signal recovery tasks.
For simulations we used cavity and waveguide parameters accepted as standard for GaAs integrated photonic circuits.
We found that such integrated photonic reservoirs show a comparable performance to other implementations found in the literature, which shows a great potential of this maximum integration-density approach.

\section{System of coupled nonlinear microcavities}

\begin{figure}[hbtp!]
	\begin{center}
		\includegraphics[width=\textwidth]{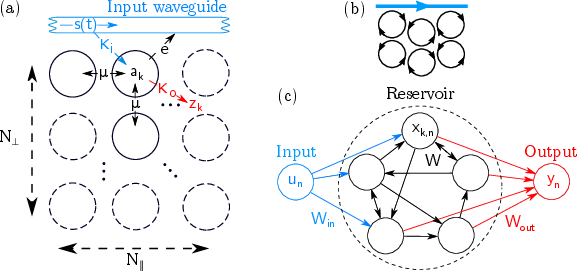}
	\end{center}
	\caption{(a)~Proposed photonic Reservoir Computer (b)~MRR mode direction matching and (c)~Echo State Network.}
	\label{fig:dims}
\end{figure}

The backbone of our photonic RC is a $N_\parallel \times N_\perp$ square grid of directly coupled nonlinear cavities~(see Figure~\ref{fig:dims}(a)).
One side of the grid is coupled to the input waveguide according to cavity-waveguide coupling coefficient $\kappa$.
Through this waveguide an optical input $u(t)$ is injected.
Assuming intra-cavity and cavity-waveguide coupling strengths are weak, the coupled-mode theory can be applied~\cite{manolatou1999}.
For a purely linear case, the electric field of microcavity modes $\bi{a}(t)$ can be described with
\begin{equation}
	\frac{\rm{d}\bi{a}}{{\rm d}t} = \left({\rm i}\bomega - \frac{\bGamma^{\rm o} + \bGamma^{\rm e}}{2} \right)\odot\bi{a}
	+ (\hat M^{\mu} + \hat M^{\kappa}) \bi{a} + \bi{K}^{\rm i} s(t),
	\label{eq:cavity-eq}
\end{equation}
where
$\bomega$ are the cavities' resonant frequencies,
$\bGamma^{\rm o}$ are intrinsic optical losses,
$\bGamma^{\rm e}$ are optical losses to waveguide,
$\odot$ is the the Hadamard product,
$\hat M^{\mu}$, $\hat M^{\kappa}$ are direct and waveguide-assisted cavity-cavity coupling matrices,
$\bi{K}^{\rm i}$ is an input waveguide-cavity coupling matrix (see~\ref{sec:wg-cav-matrix}).
Here, $\hat M^{\mu}_{km} = -(\hat M^{\mu}_{mk})^* = |\mu_{km}| \exp({\rm i}\varphi_{km})$~\cite{haus1984}~(Section 7.5) if the $k$-th and $m$-th cavities are directly coupled
and $K^{{\rm i}}_k = \kappa^{{\rm i}}_{k}\exp({\rm i} \varphi^{{\rm i}}_{k})$, with $k,m \in 1 \dots N$, where $N = N_\parallel N_\perp$.
For simplicity, we assume that $\Gamma^{{\rm o}}_k = \Gamma^{{\rm o}}$, $|\mu_{km}| = \mu$ and $\kappa^{{\rm i}}_{k} = \kappa^{{\rm i}}$, if the $k$-th cavity is coupled to the waveguide.
While this is a good approximation, $\Gamma^{{\rm o}}_k$, $|\mu_{km}|$ and $\kappa^{{\rm i}}_{k}$ can vary to some degree, e.g. due to fabrication tolerances.
The phase terms $\varphi_{km}$ and $\varphi^{{\rm i}}_{k}$ are defined by a chip geometry and are considered independent and identically distributed (i.i.d.).
The strength of intra-cavity and the cavity-waveguide coupling is weak when fulfilling
\begin{equation}
	|{\rm i}\bomega\odot\bi{a}| \gg \left|-\frac{1}{2}\bGamma^{\rm e}\odot\bi{a}
	+ (\hat M^{\mu} + \hat M^{\kappa}) \bi{a} + \bi{K}^{\rm i} s(t)\right|.
\end{equation}

Direct coupling can only induce an interaction between cavities when resonances sufficiently aligned.
For that reason, unless stated otherwise, we assume $\omega_k = \omega_{{\rm 0}}$.
Moreover, the direction of cavity modes should be taken into account.
For instance, each microring resonator (MRR) resonance is double-degenerate with counter-propagating modes.
If a MRR-based reservoir is implemented exactly as in Figure~\ref{fig:dims}(a), both modes of each cavity can be excited and $\bi{a}$ would have $2N$ components.
In photonic crystal (PhC) cavities, however, resonances are nondegenerate and therefore $\bi{a}$ has $N$ components.
For simplicity we transform the MRR reservoir such that only one mode of each MRR cavity is excited (see Figure~\ref{fig:dims}(b)).

In a fully integrated system, cavities would also be coupled to readout waveguides.
This, however, can be challenging to realize for cavities inside the grid, since the compact cavity arrangement leaves no space for such waveguides.
One could consider using $N_\parallel\times 1$ or $2\times N_\perp$ grids, where all cavities are accessible.
An alternative could be the use of multiple photonic layers~\cite{sacher2018} or three-dimensional waveguides~\cite{moughames2020}.
Output signals of these waveguides are
$
	\bi{z}(t) = \bi{K}^{\rm o}\odot\bi{a}(t),
$
where $\bi{K}^{\rm o}$ is a vector of output waveguide coupling coefficients, for which we assume an uniformity according to $K^{{\rm o}}_k = \kappa^{{\rm o}}$.
These waveguides introduce an additional optical loss, which has to be included in $\bGamma^{\rm e}$.

We consider three nonlinear effects commonly present in integrated photonic microresonators: TPA, FCD and the Kerr effect.
TPA can be described by adding
\begin{equation}
	-\frac{\Gamma^{{\rm TPA}}(a_{k})}{2}a_{k}
	= -\frac{1}{2}\frac{\beta_2 c^2}{n^2 V^{{\rm TPA}}}|a_{k}|^2 a_{k}
	\label{eq:tpa}
\end{equation}
to the right side of Eq.~\ref{eq:cavity-eq}~\cite{uesugi2006}.
Here, $\beta_2$ is a TPA coefficient, $c$ the speed of light, $n$ the refractive index and $V^{\rm TPA}$ is a nonlinear effective volume~\cite{moille2016}
\begin{equation}
	V^{{\rm TPA}} = \left( \frac{\varepsilon_0^2}{4}\int \varepsilon_{\rm r} \chi_{\rm r} \frac{2|\bi{U}|^4 + |\bi{U}\cdot\bi{U}|^2}{3}dV \right)^{-1},
\end{equation}
where $\varepsilon_{\rm r} = \varepsilon(\bi{r})/\varepsilon_{\rm max}$,
$\chi_{\rm r} = \chi^{(3)}(\bi{r}) / \chi^{(3)}_{\rm max}$
and $\bi{U}(\bi{r})$ is a cavity's normal mode, normalized such that
$
0.5\int \varepsilon_{\rm 0}\varepsilon_{\rm r} |\bi{U}(\bi{r})|^2 d\bi{r} = 1.
$

FCD causes a microcavity's resonance to shift by $\Delta \omega_k$ as a function of free electron density $N_k$~\cite{bennett1990}.
This effect adds a term
$
	{\rm i}({\rm d}\omega/{\rm d}N) N_{k} a_{k}
	$
to the right side of Eq.~\ref{eq:cavity-eq}~\cite{xu2006}.
Such free electrons could be generated via a linear photon absorption or TPA.
The former is possible in direct-bandgap semiconductors, but causes high optical loss independent of the input power, unlike with TPA, where it linearly increases with the optical energy stored in a cavity.
Considerable losses in systems without gain make large networks unrealistic.
We therefore exclude linear photon absorption as sustainable effect producing nonlinearity and only consider TPA for the free electron generation.
The density of TPA-generated electrons is computed by~\cite{uesugi2006}
\begin{equation}
	\frac{{\rm d} N_{k}}{{\rm d}t} = -\Gamma^{{\rm c}} N_{k} + \frac{\Gamma^{{\rm TPA}}(a_{k})}{2\hbar \omega V^{\rm{c}}}I_{k},
	\label{eq:tpa_n}
\end{equation}
where $\Gamma^{{\rm c}}$ is a free electron recombination rate, $V^{\rm{c}}$ is a volume across which electrons spread through a diffusion, and $I_{k} = |a_{k}|^2$ is the optical intensity inside the $k$-th cavity.

The Kerr effect causes a resonance frequency shift in response to a stronger electric field inside a cavity and is described by adding
$
	{\rm i}({\rm d}\omega/{\rm d}I) I_{k}a_k
	\label{eq:kerr}
$
to the right side of Eq.~\ref{eq:cavity-eq}.
While the impact of the Kerr effect is typically weaker than FCD, it is still useful to be considered as an isolated case, as, unlike FCD with TPA, it provides a resonance frequency shift without an additional optical losses.

\section{A Photonic Reservoir Computer based on directly coupled nonlinear cavities}
\label{sec:photonic-reservoir}
A RC is often introduced as an Echo State Network~(ESN)~\cite{lukovsevivcius2012}~(see Figure~\ref{fig:dims}(c)):
\begin{equation}
	\bi{x}_{n+1} = (1-\alpha)\bi{x}_n + \alpha\tanh(\hat W \bi{x}_{n} + \hat W^{{\rm in}} [1; \bi{u}_{n+1}]),
	\label{eq:reservoir-eq}
\end{equation}
where
$\bi{x}_n$ and $\bi{u}_n$ are vectors of neuron activations and inputs at a time step $n \in [1,2,\dots{T}]$,
$\alpha$ is a leaking rate,
$\hat W$ and $\hat W^{{\rm in}}$ are a recurrent and an input weight matrices.
The ESN's output is defined as
$
\bi{y}_n = \hat W^{{\rm out}} \bi{x}_n,
$
where $\hat W^{{\rm out}}$ is an output weight matrix.
The readout training for software-based models is typically performed using ridge regression~\cite{lukovsevivcius2012}, and here we use the same approach for our simulation.
Assume an ESN is being trained to solve a task with input signals $\bi{u}_n$ and the ideal target outputs $\bi{y}^{\rm tgt}_{n}$.
First, compute update equations for $T$ steps to obtain a set of ESN state vectors $\bi{x}_n$, which are consequently appended to create a matrix $\hat X$.
Similarly, from $\bi{y}^{\rm tgt}_{n}$ one creates $\hat Y^{\rm tgt}$.
The optimal output weight matrix is then given by
\begin{equation}
	\hat W^{{\rm out}} = \hat Y^{\rm tgt} \hat X^{\rm T} \left(\hat X \hat X^{\rm T} + \zeta \langle |X_{ij}|^2 \rangle_{ij} \hat I\right)^{-1},
	\label{eq:ridge}
\end{equation}
where $\langle \dots \rangle_{ij}$ is an average over all matrix indices $ij$, $\zeta$ is a regularization constant and $\hat I$ is the identity matrix, superscript $(\dots)^{\rm T}$ is the matrix transpose operation.

In the previous section, we proposed a photonic system that receives optical signal $s(t)$ as an input and provides a set of optical signals $\bi{z}(t)$.
We suggest to treat this system as a reservoir computer.
In the integrated photonic hardware, readout weights could be realized based on tunable attenuators with transmission $w_{k}$, phase-shifters according to $\varphi_k$ and a combiner tree to implement a linear combination of $z_k(t)$ that creates the output optical signal $y(t)$:
\begin{equation}
	y(t) = \sum\limits_k w_k \exp({\rm i}\varphi_k) z_k(t) = \bi{W^{{\rm out}}} \bi{z}(t).
\end{equation}
This way, the conversion of optical input $s(t)$ to optical output $y(t)$ is done fully optically, as electronics are only controlling attenuators and phase shifters that are constant during a computation.


\section{Linear regime}
\label{sec:dimensionality-linear}

We first consider the case when the input signal is not sufficiently strong to induce a nonlinear response by the microcavities.
An important value for the following discussion is the input signal's bandwidth ${\rm BW}$ that we define as the span between its spectral half-power positions.

\begin{figure}[hbtp!]
	\begin{center}
		\includegraphics[width=\textwidth]{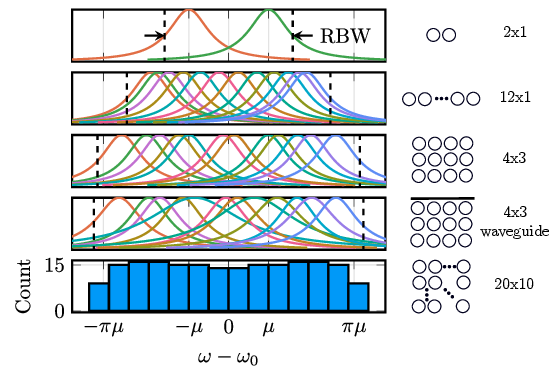}
	\end{center}
	\caption{
		Supermodes of directly coupled microcavities for various grid geometries.
		Bottom figure shows a histogram of supermode frequencies.
	}
	\label{fig:splitting}
\end{figure}

As we will show, computational properties of such photonic reservoirs depend mostly on the microcavities' spectral characteristics.
In that regard, an important consequence when directly coupling microcavities is a formation of supermodes with split resonances (see Figure~\ref{fig:splitting}).
Consider the linear part of Eq.~\ref{eq:cavity-eq}:
\begin{equation}
	\rm{diag}\left({\rm i}\bomega - \frac{\bGamma^{\rm o} + \bGamma^{\rm e}}{2}\right) + \hat M^{\mu} + \hat M^{\kappa},
	\label{eq:eigenmatrix}
\end{equation}
eigenvalues of which are $\lambda^{\rm E}_k$.
Then the average supermode bandwidth is $\Gamma^{{\rm S}}/2\pi = \left\langle -2{\rm Re}\left(\lambda^{\rm E}_k\right) \right\rangle_k/2\pi$.
Importantly, the coupling-induced supermode splitting effectively increases the reservoir's response bandwidth significantly beyond the bandwidth of a single resonator.
We define the reservoir bandwidth~(RBW) as the maximum distance between the network's supermodes measured at half intensity.
For example, without the input waveguide, two identical independent cavities with resonances frequencies at $\omega_{{\rm 0}}$ and coupled with rate $\mu$ become split $\omega_{{\rm 0}} \pm |\mu|$, leading to ${\rm RBW} \approx (2|\mu|+\Gamma^{{\rm S}})/2\pi$.
For multiple indentical cavities in a chain, Eq.~\ref{eq:eigenmatrix} is a tridiagonal Toeplitz matrix, eigenvalues of which are given by~\cite{kulkarni1999}
\begin{equation}
	\lambda^{\rm E}_k = \left({\rm i}\omega_0 + \frac{\Gamma^{\rm o}}{2}\right) + 2\sqrt{-|\mu|^2}\cos\left(\frac{k\pi}{N+1}\right), ~~~~k = 1\dots N,
\end{equation}
meaning that ${\rm RBW} \approx (4|\mu| + \Gamma^{{\rm S}}) / 2\pi$.
For non-trivial connectivity architectures an analytical derivation of ${\rm RBW}$ becomes non-trivial.
However, in numerical simulations we found that supermode resonances stay within a $\omega_{{\rm 0}} \pm \pi|\mu|$ interval~(see Figure~\ref{fig:splitting}).
We therefore assume that ${\rm RBW} \approx (2\pi|\mu|+\Gamma^{{\rm S}})/2\pi$ for such directly coupled microcavity system.
Furthermore, as will be shown later, in the cases we are interested in $\Gamma^{{\rm S}} \ll |\mu|$, and hence ${\rm RBW} \approx |\mu|$.

An important computing metric of reservoirs is dimensionality $D$, which corresponds to the systems' degrees of freedom.
Since reservoirs, and ANNs in general, perform a high-dimensional expansion of input signals~\cite{lukovsevivcius2012}, a higher dimensionality usually leads to a better computing capacity.
To determine the dimensionality we generate a random sequence $\tilde s_0(t_n) = \sqrt{I_0}(t_n)\exp({\rm i}\varphi_0(t_n))$, where $\sqrt{I^0}(t_n)$ and $\varphi_0(t_n)$ are randomly sampled according to a white noise distribution.
Then, an 8th-order Butterworth low-pass filter with a given bandwidth ${\rm BW}$ is applied on $\tilde s_0(t_n)$ to obtain $s_0(t_n)$, which is then linearly interpolated to $s_0(t)$.
This signal is used to modulate an optical carrier that is injected into the reservoir as an input.
Here and throughout the article the integration of differential equations is carried out using the {\it DifferentialEquations.jl} library~\cite{rackauckas2017} in the Julia programming language~\cite{bezanson2017}.
The photonic reservoir's dimensionality is then determined by applying the Principal Component Analysis~(PCA)~\cite{pearson1901} on $\bi{z}_n$.
Importantly, for the PCA we split electric fields into a real and an imaginary part, which implies that the maximum dimensionality our reservoir can have is $2N$.
It is therefore convenient to introduce a normalized dimensionality $\tilde D = D / 2N$.

At this stage, it is beneficial to consider the supermode concept in more detail.
A supermode is an independent, i.e. an orthogonal oscillation of the coupled system's electric fields.
Each supermode can be considered as a virtual microcavity, that performs a pass-band filtering at its resonant frequency and bandwidth.
Crucially, the coupling-induced splitting of supermode resonances makes their frequencies vary, promising high dimensionality.

\begin{figure}[hbtp!]
	\centering
	\includegraphics[width=\textwidth]{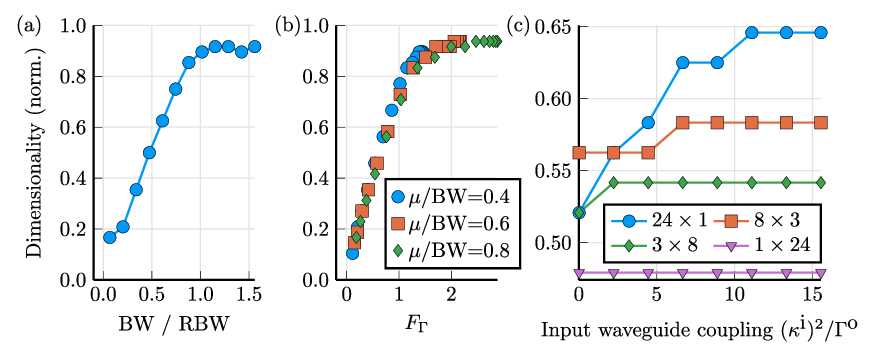}
	\caption{
		Dimensionality of linear photonic reservoir w.r.t. (a)~input signal bandwidth, (b)~average supermode spacing and (c)~input waveguide coupling strength.
		Reservoir parameters are given in Table~\ref{tab:mg} and by default ${\rm BW} > {\rm RBW}$, but in (a)~$\mu$ increased to 80~GHz to increase upper limit of dimensionality, in (b) each point corresponds to specific $\mu$ and $\Gamma^{{\rm o}}$.
		(c)~different grid geometries, see legend.
	}
	\label{fig:dim-bundle}
\end{figure}

To start, a supermode needs to receive an input.
For that, we consider the impact of the injection BW on the reservoir's dimensionality.
As we see in Figure~\ref{fig:dim-bundle}(a), when ${\rm BW} \ll {\rm RBW}$ the dimensionality is low.
In this case, a large number of supermodes are off-resonance relative to the injected field, and hence they cannot be excited.
With an increasing BW, more and more supermodes are able to interact with the injected information.
As a consequence, the reservoir's dimensionality increases until ${\rm BW} = {\rm RBW}$, when all supermodes become excited.
Accordingly, the physics of supermodes in coupled arrays result in a necessary bandwidth matching condition for maximizing the system's dimensionality:
$
	{\rm BW} \geq {\rm RBW}.
$
Importantly, here the network was driven with white noise, i.e. by a signal with a flat spectrum.
However, in realistic and application-relevant situations such a signal is unlikely.
Certain supermodes will be excited less than others and accordingly contribute less to the networks dimensionality.

The microcavity network's dimensionality relies upon individual supermodes being driven by different aspects of an input signal, which requires supermodes to be sufficiently separated.
Therefore, to quantify this metric, we consider an average supermode spacing ${\rm RBW} / N$, normalized by the average supermode bandwidth
\begin{equation}
	F_\Gamma = \frac{{\rm RBW}/N}{\Gamma^{{\rm S}}/2\pi},
\end{equation}
which can be considered as a measure of spectral span available for individual supermodes.
%
In Figure~\ref{fig:dim-bundle}(b), we see that the dimensionality is low for $F_\Gamma \ll 1$ and for $F_\Gamma > 1$ dimensionality is maximized.
This threshold may change depending on system parameters, however, we find that it is typically comparable to unity.
Therefore a second requirement for a high dimensionality is
$
	F_\Gamma \gtrapprox 1.
$
Again, this is not a sufficient condition, since it does not ensure that all supermodes are separated.
Spacing between supermodes is typically inhomogeneous, and supermodes can potentially overlap, even if $F_\Gamma$ is large.
This effect can be mitigated through a stronger coupling to the input waveguide $\kappa^{{\rm i}}$, which can shift resonances or increase the bandwidth of some supermodes, as can be seen in Figure~\ref{fig:splitting} for a $4 \times 3$ reservoir with and without the input waveguide.
In Figure~\ref{fig:dim-bundle}(c), we see that this approach can be effective when $N_\perp$ is low, i.e. when overall coupling to the input waveguide is relatively stronger for the supermode splitting.

\begin{figure}[hbtp!]
	\centering
	\includegraphics[width=\textwidth]{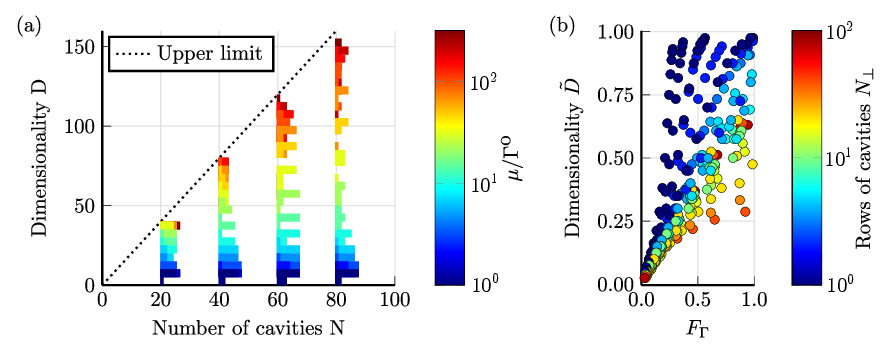}
	\caption{
		(a)~Scalability of linear reservoir dimensionality w.r.t. optical losses.
		Dimensionalities of all possible combinations of $N_\parallel $ and $N_\perp$ for given $N$ are shown as histograms.
		Reservoir parameters are shown in Table~\ref{tab:mg}, with $\mu = 80$~GHz, reduced $k_{\rm o}$ and ${\rm BW} > {\rm RBW}$.
		(b)~Impact of $F_\Gamma$ and grid geometry on dimensionality.
		Both figures use the same dataset.
	}
	\label{fig:dim-vs-q-n}
\end{figure}

From $F_\Gamma$ a scaling behaviour of dimensionality can be determined.
Since $F_\Gamma \propto N^{-1}$, a larger number of cavities leads to a lower $F_\Gamma$, which leads to a lower $\tilde D$.
This, however, can be compensated by lowering optical losses, as $F_\Gamma \propto (\Gamma^{{\rm S}})^{-1}$.
Indeed, in Figure~\ref{fig:dim-vs-q-n}(a) we see that with a high $\mu/\Gamma^{{\rm o}}$ (which is analogous to $F_{\Gamma}$), the dimensionality can reach the theoretical limit.
Furthermore, dimensionality varies for a given number of cavities depending on their geometric arrangement.
A minor fluctuation of dimensionality is due to a seemingly random nature of supermode formation -- supermode frequencies of a $24 \times 1$ grid are different from a $12 \times 2$ grid.
However, there are also outliers with a considerably lower dimensionality.
In these cases $N_\perp$ is high, which allows supermodes to localize spatially further from the input waveguide.
As a result, these supermodes become too weakly coupled to the input and as a consequence contribute less to the system's dimensionality~(see Figure \ref{fig:dim-vs-q-n}(b)).
In addition to that, a lower $N_\perp$ allows for a higher dimensionality even at a low $F_\Gamma$, likely due to the impact of input waveguide discussed above.

\begin{table}[hbtp!]
	\caption{
		Default reservoir parameters.
		Most parameters are omitted as they only impact the required input power.
	}
	\label{tab:mg}
{\small
	\begin{center}
		\begin{tabular}[c]{r|r|l|l}
			\multicolumn{1}{c|}{Parameter} &
			\multicolumn{1}{c|}{Value} &
			\multicolumn{1}{c|}{Units} &
			\multicolumn{1}{c}{Comments} \\
			\hline
			Geometry & $8 \times 3$ grid & & GaAs MRRs\\
			$\mu$ & $25$ & GHz & defines ${\rm RBW}$\\
			$\Gamma^{{\rm c}}$ & $28.5$ & GHz & \cite{ibrahim2003}\\
			Q-factor & $3\cdot 10^5$ & & provide a near-optimal $\Gamma^{{\rm S}}/2\pi \approx 0.9$ GHz $\approx {\rm RBW}/N$\\
			$k_{{\rm i}}^2$ & $4\pi$ & GHz & every other cavity next to input waveguide is coupled\\
			$k_{{\rm o}}^2$ & $\pi$  & GHz & provides sufficiently strong $z_k(t)$
		\end{tabular}
	\end{center}
}
\end{table}

\section{Nonlinear regime}
\label{sec:dimensionality-nonlinear}

In order to induce a nonlinear response from the photonic system, an optical power of at least $P_{{\rm NL}}$ needs to be injected into the cavities.
This power depends on material, cavity parameters and which particular nonlinearity is to be leveraged.
We define $P_{{\rm NL}}$ as an input power at which a nonlinear effect becomes equal to a related linear effect.
Two-photon absorption increases losses inside the cavities according to $\Gamma^{{\rm TPA}}$, and it is natural to place $P_{{\rm NL}}$ where these nonlinear losses start matching linear optical losses, i.e. $\Gamma^{{\rm TPA}} \approx \Gamma^{{\rm S}}$.
FCD and the Kerr effect cause resonance shift $\Delta\omega$, which, in turn, is natural to be compared to an average bandwidth of supermodes, hence also $\Gamma^{{\rm S}}$.
Therefore, an optical loss rate is the normalization factor determining the cavity array's nonlinearity power threshold for all considered nonlinear effects.

For a single cavity injected with a monochromatic wave one can determine $P_{{\rm NL}}$ analytically (see supplementary material).
The result is shown in Figure~\ref{fig:refpower}(a) for various integrated platforms for MRR and PhC cavities based on a data obtained from literature (see~Table~\ref{tab:refpower}), and the horizontal axis shows an operating frequency that is ${\rm BW}/2$.
Assume a cavity with a bandwidth $\Gamma^{{\rm S}}$ and an electron relaxation rate $\Gamma^{{\rm c}}$.
This cavity can receive most optical power of an injected signal if ${\rm BW} \le \Gamma^{{\rm S}}/2\pi$.
However, a low $\Gamma^{{\rm S}}$ corresponds to higher loaded Q-factor and allows for a stronger nonlinearity and, consequently, a lower $P_{{\rm NL}}$.
We therefore assume that ${\rm BW} \approx \Gamma^{{\rm S}}/2\pi$.
Additionally the scale of ${\rm BW}$ relative to $\Gamma^{{\rm c}}$ is also important.
In the case when ${\rm BW} \ll \Gamma^{{\rm c}}$, the generated free electrons recombine too fast and FCD becomes negligible, whereas TPA becomes the dominant nonlinearity.
Then, $P_{{\rm NL}}$ is computed w.r.t. $\Gamma^{{\rm TPA}}$, see dashed line in~Figure~\ref{fig:refpower}(a).
When ${\rm BW}$ is comparable to $\Gamma^{{\rm c}}$ FCD becomes stronger than TPA.
Then, $P_{{\rm NL}}$ is computed w.r.t. $\Delta\omega$ and this case is shown as a solid line in~Figure~\ref{fig:refpower}(a).
Finally, when ${\rm BW} \gg \Gamma^{{\rm c}}$, the electron density can't react to the changes to the input signal in time and FCD becomes irrelevant, see dotted line in~Figure~\ref{fig:refpower}(a).
%
%
Compared to MRRs, PhC cavities require less power and allow a faster operation on the same material, as a smaller mode volume corresponds to a stronger light confinement and a stronger nonlinearity, while a higher surface area enhances $\Gamma_{\rm c}$~\cite{tanabe2005}.

For a network comprised of a large number of cavities, analytical treatment is not immediately tractable.
We therefore resort to a numerical approach.
Consider a reservoir with TPA as a nonlinearity.
Since its supermodes are split due to direct coupling and, as we found in the previous section, $\Gamma^{\rm S} \ll {\rm RBW}$, a low-bandwidth excitation of this nonlinearity is not suitable.
Therefore, the input changes in time and modulates $\Gamma^{{\rm TPA}}_{k}$ on a potentially individual resonator level.
In order to then compare it to the scalar $\Gamma^{{\rm S}}$, we consider its standard deviation across many input samples $\sigma_t[\Gamma^{{\rm TPA}}_{k}]$.
Another natural option could be an average over samples, however, consider a case when TPA is dominant and
$\Gamma^{{\rm TPA}}_{k}(a_k(t)) = \langle\Gamma^{{\rm TPA}}_{k}\rangle_t + \delta\Gamma^{{\rm TPA}}_{k}(a_k(t))$ with $\delta\Gamma^{{\rm TPA}}_{k}(a_k(t)) \ll \langle\Gamma^{{\rm TPA}}_{k}\rangle_t \approx \Gamma^{{\rm S}}$,
then the loss terms in Eq.~\ref{eq:cavity-eq} for the $k$-th cavity is
\begin{equation}
	\frac{{\rm d}a_k}{{\rm d}t}
	= \left(-\frac{\Gamma^{{\rm o}}_k - \Gamma^{{\rm e}}_k}{2} - \langle\Gamma^{{\rm TPA}}_{k}\rangle_t - \delta\Gamma^{{\rm TPA}}_{k}(a_k(t)) \right) a_k
	\approx \left(- \frac{\Gamma^{{\rm o}}_k - \Gamma^{{\rm e}}_k}{2} - \langle\Gamma^{{\rm TPA}}_{k}\rangle_t \right) a_k.
\end{equation}
Here, $\langle\Gamma^{{\rm TPA}}_{k}\rangle_t$ effectively increases the linear loss and hence the $k$-th cavity is close to linear.
However, averaging would still classify the cavity as a nonlinear, since $\langle\Gamma^{{\rm TPA}}_{k}(a_k(t))\rangle_t \approx \langle\Gamma^{{\rm TPA}}_{k}\rangle_t \approx \Gamma^{{\rm S}}$, whereas with the standard deviation $\sigma_t\left[\Gamma^{{\rm TPA}}_{k}(a_k(t))\right] = \sigma_t\left[\delta\Gamma^{{\rm TPA}}_{k}(a_k(t))\right] \ll \Gamma^{{\rm S}}$.
We then compute $\langle \sigma_t[\Gamma^{{\rm TPA}}_{k}]\rangle_k$ to obtain a scalar TPA rate for a given input power $P$.
Repeating this process for a range of $P$ we obtain a largely monotonic curve $P(\langle \sigma_t[\Gamma^{{\rm TPA}}_{k}]\rangle_k)$, through which we can determine the input power necessary to induce a particular level of TPA-induced nonlinearity.
In that case $P_{{\rm NL}} = P(\Gamma^{{\rm S}})$.
For FCD and the Kerr effect the process to obtain a scalar characteristic capturing the network's relevant response is identical, except that $\Delta\omega_k$ is used instead of $\Gamma^{{\rm TPA}}_{k}$.
In Figure~\ref{fig:refpower}(b) we see that $P_{{\rm NL}}$ scales linearly with the number of cavities, provided $N_\perp$ is limited.
Indeed, for large $N_\perp$ some supermodes can localize far from the input waveguide and couple to input much weaker than those near it.
This creates a nonlinearity imbalance in the reservoir and to reach $\langle \sigma_t[\Gamma^{{\rm TPA}}_{k}]\rangle_k = \Gamma^{{\rm S}}$, more optical power is needed.


\begin{figure}[hbtp!]
	\begin{center}
		\includegraphics[width=\textwidth]{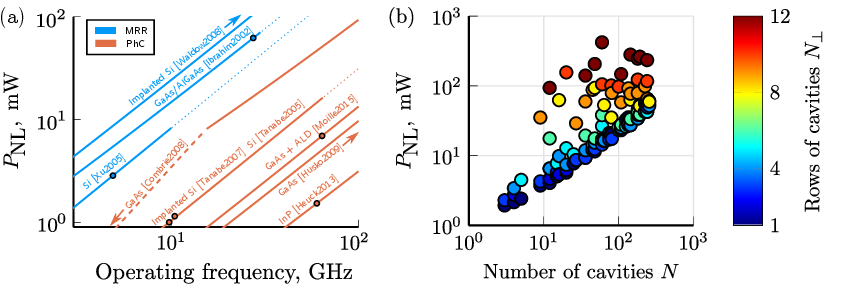}
	\end{center}
	\caption{
		(a)~Comparison of material platforms w.r.t. $P_{{\rm NL}}$ and an operating frequency of a single MRR or PhC cavity, details in text.
		Here, ALD is atomic layer deposition.
		For references see Table~\ref{tab:refpower}.
		(b)~Scaling of $P_{{\rm NL}}$ w.r.t. the number of cavities.
		Reservoir parameters are given in Table~\ref{tab:mg}, BW matches RBW.
	}
	\label{fig:refpower}
\end{figure}
\begin{table}[hbtp!]
	\caption{
		References used in Figure~\ref{fig:refpower}.
	}
	\label{tab:refpower}
	\begin{center}
		\begin{tabular}[c]{l l l}
			\hline
			\multicolumn{1}{c}{Key} &
			\multicolumn{1}{c}{Cavity} &
			\multicolumn{1}{c}{Reference} \\
			\hline
			\multicolumn{2}{l}{\textbf{GaAs}} \\
			Ibrahim2002 & MRR & \cite{ibrahim2002} \\
			Combrie2008 & L$_5$ PhC & \cite{combrie2008} \\
			Husko2009 & H$_0$ PhC & \cite{husko2009} \\
			Moille2015 & 2$\times$ H$_0$ PhC + ALD & \cite{moille2015} \\
			\multicolumn{2}{l}{\textbf{Silicon}} \\
			Xu2005 & MRR & \cite{xu2005} \\
			Tanabe2005 & L$_3$ and L$_4$ PhC & \cite{tanabe2005} \\
			\multicolumn{2}{l}{\textbf{Implanted silicon}} \\
			Waldow2008 & MRR & \cite{waldow2008} \\
			Tanabe2007 & H$_1$ PhC & \cite{tanabe2007} \\
			\multicolumn{2}{l}{\textbf{InP}} \\
			Ibrahim2003 & MRR & \cite{ibrahim2003} \\
			Heuck2013 & H$_0$ PhC & \cite{heuck2013} \\
		\end{tabular}
	\end{center}
\end{table}

Nonlinearity can result in a reservoir operating in a chaotic regime, which for most purposes is harmful to computation.
Chaos corresponds to the situation where a network's state is sensitive to infinitesimal changes in the input, and hence the unavoidable presence of noise in hardware would render computations not reproducible for even perfectly identical inputs.
For characterizing this effect we carry out a consistency analysis~\cite{skalli2022}.
Consistency is a property describing the reproducibility of a dynamical system's (here the reservoir's) responses under an injection of slightly different input signals~\cite{uchida2004}.
%
We generate a set of inputs in the form $s_n(t) = s_0(t) + \Delta s_n(t)$ where $s_0(t)$ and $\Delta s_n(t)$ are values sampled from a complex-valued white noise distribution, which filtered with a Butterworth 8th order pass-band filter.
$s_0(t)$ is the ``true input'', while $\Delta s_n(t)$ emulates an impact of noise in an experiment and we attain a signal to noise ratio of 20~dB by scaling the amplitude of $\Delta s_n(t)$.
Each $s_n(t)$ is injected into a separate, but identical reservoir.
The absolute value of correlation between trajectories of $k$-th cavity of $i$-th and $j$-th reservoirs $z_{k}^{i}(t)$ and $z_{k}^{j}(t)$ is then computed~\cite{park2018}
\begin{equation}
	\gamma^{ij}_{k}
	= \left| \frac{\mathrm{E}\left[z^{i}_{k} (z^{j}_{k})^*\right] - \mathrm{E}\Big[z^{i}_{k}\Big]\mathrm{E}\left[(z^{j}_{k})^*\right]}{\sqrt{\mathrm{E}\Big[|z^{i}_{k}|^2\Big]}\sqrt{\mathrm{E}\left[|z^{j}_{k}|^2\right]}} \right|,
\end{equation}
where $\mathrm{E}[A]$ is expected value of $A$.
The consistency for this pair of reservoirs is a root mean square (RMS) across all their cavities
$
	\gamma^{ij} = \sqrt{\langle (\gamma^{ij}_{k})^2 \rangle_{k}}
$
and equally the consistency of system $\gamma$ is the RMS of $\gamma^{ij}$ of all combinations of $i$ and $j$.
It is typically desirable that the consistency is high~\cite{lymburn2019}.

\begin{figure}[hbtp!]
	\centering
	\includegraphics[width=\textwidth]{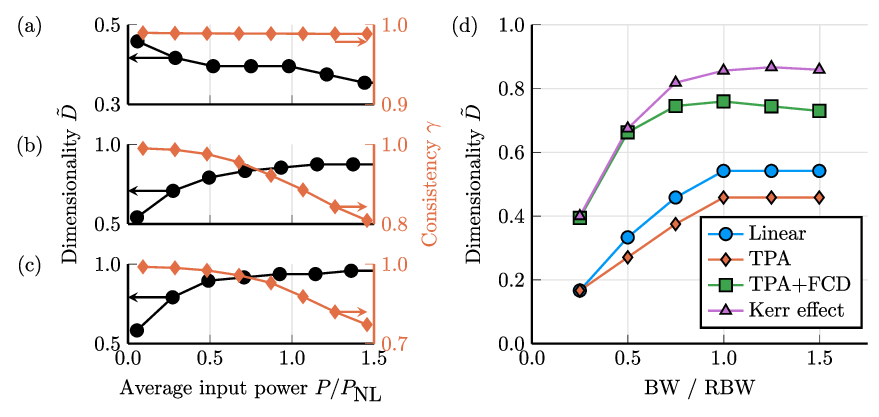}
	\caption{
		Effect of (a) TPA (b) TPA and FCD (c) the Kerr effect on reservoir dimensionality and consistency.
		Reservoir parameters are given in Table~\ref{tab:mg}.
		(d) each point corresponds to an input power at which $\gamma = 0.98$.
	}
	\label{fig:dim-cons}
\end{figure}

In Figure~\ref{fig:dim-cons} we show the consistency and dimensionality of reservoirs with different nonlinearities as a function of input power.
Noteworthy, we find that a stronger TPA reduces the dimensionality, while the consistency remains constant and close to unity.
The drop of dimensionality is a consequence of the $F_{\Gamma}$ reduction, as increasing TPA-induced losses effectively increase $\Gamma^{{\rm S}}$.
A high consistency could be explained by the negative-feedback nature of TPA, where the system's response to a stronger input is an increased damping that often stabilizes the system.
According to the data in Figure~\ref{fig:dim-cons} FCD and the Kerr effect increase the dimensionality, but eventually the system loses consistency for too high input power.
Generally, consistency decreases when $\langle \sigma_t[\Delta\omega_k] \rangle_k$ approaches $\Gamma^{{\rm S}}$, i.e. when input power surpasses $P_{{\rm NL}}$.
The effect causing the dimensionality increase is the nonlinear shift of individual supermodes, which otherwise might overlap in the linear regime.
We found that generally other system parameters have a relatively small impact compared to $P_{{\rm NL}}$.

Interestingly, the nonlinear shift of supermode frequencies also compensates to some degree the negative impact on dimensionality caused by a mismatch between ${\rm BW}$ and ${\rm RBW}$ (see Figure~\ref{fig:dim-cons}(d)).
This could be explained by the fact that a cavity might belong to multiple supermodes.
When a cavity resonance shifts, it causes multiple supermodes to shift as well, even those that are not excited in a linear regime.
If the shift is strong enough, these supermodes can end up inside the input bandwidth and receive input.
Moreover, as ${\rm BW}$ reduces, a stronger nonlinearity is needed to move resonances into the input bandwidth, and at some point the concept of supermodes stops being applicable.

To conclude, nonlinearity in general has a non-trivial effect on a reservoir.
Importantly, this sensitively depends on the particular nonlinear effect that is leveraged.
On one hand, dimensionality is reduced by TPA, but increased by FCD and the Kerr effect.
On the other hand, using TPA the reservoir stays consistent, while FCD and the Kerr effect are harmful to consistency.
However, in the case of FCD and Kerr nonlinearities, the increase of dimensionality comes before the loss of consistency.
Hence, there is an interval of optimal input power that allows for a better dimensionality with a high consistency.

\section{Computing performance}
We now turn to evaluating the performance when applying a direct coupled cavity array reservoir to computing tests.
Here, we consider the Mackey-Glass prediction task, a commonly used benchmark in the RC literature, and the optical signal recovery task that demonstrates a relevant practical application of such a system~\cite{argyris2022}.

\subsection{Mackey-Glass prediction task}
\label{sec:mg}

The Mackey-Glass equation is a first-order time-delayed differential equation:
\begin{equation}
	\frac{{\rm d}\xi}{{\rm d}t} = \frac{\alpha \xi(t-\tau)}{1 + \xi(t-\tau)^g} - \gamma \xi(t),
	\label{eq:mackey-glass}
\end{equation}
where we choose $\tau = 17$, $\alpha = 0.2$, $g = 10$ and $\gamma = 0.1$, with which $\xi(t)$ exhibits a moderately chaotic behaviour~\cite{mackey1977}.
%
Integrating Eq.~\ref{eq:mackey-glass} with the Euler method with a timestep of 0.17 and downsampling the result with a ratio of $3/0.17$, we obtain a discrete series $\xi(t_n)$. 
The task is to predict $y_{n}^{\rm tgt} = \xi(t_{n+\delta})$ with $\xi(t_n)$ as an input and a given $\delta > 0$.
Both, memory and nonlinearity are required for a successful prediction, which made this task a popular RC benchmark.
The prediction performance is measured with the Normalized Root Mean Square Error~(NRMSE)~\cite{lukovsevivcius2012}:
\begin{equation}
	{\rm NRMSE}\left(\bi{y}, \bi{y}^{\rm tgt}\right) = \frac{1}{\sigma_n[y^{\rm tgt}_n]}\sqrt{\frac{1}{T}\sum\limits_{n = 1}^{T} \left|y_n - y^{\rm tgt}_{n}\right|^2}.
	\label{eq:nrmse}
\end{equation}

\begin{figure}[hbtp!]
	\centering
	\includegraphics[width=0.5\textwidth]{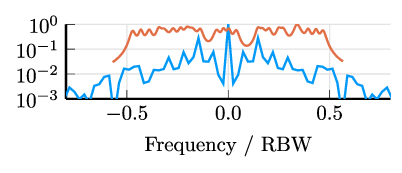}
	\caption{
		The normalized Fourier transform of input signal (blue) with reservoir resonances overlaid (orange).
		Here ${\rm BW} < {\rm RBW}$, which goes against the conclusions of Section~\ref{sec:dimensionality-linear}.
		High peaks in the input spectra correspond to a dominating harmonic ($\xi(t)$ largely oscillates, see in Figure~\ref{fig:mg-bundle}(c)), however, for a close prediction, it's important to also capture high-frequency irregularities, despite their weak amplitude.
	}
	\label{fig:mg-fourier}
\end{figure}

\begin{figure}[hbtp!]
	\centering
	\includegraphics[width=\textwidth]{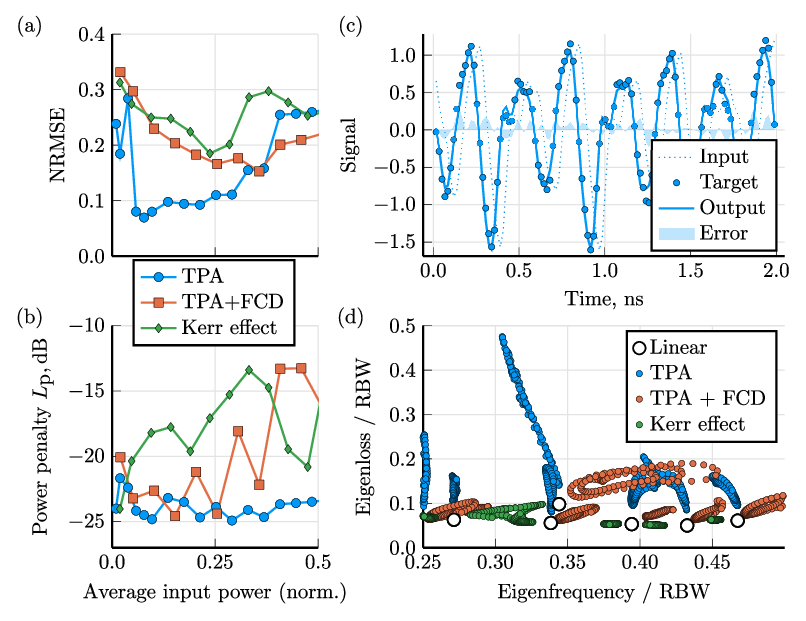}
	\caption{
		Impact of nonlinearities on (a)~prediction error and (b)~power penalty on a 3-step Mackey-Glass prediction.
		For all cases GaAs MRR parameters are used (see Table~\ref{tab:mg}), but for TPA-only ${\rm d}\omega/{\rm d}N = 0$, while for Kerr effect $\beta_2 = 0$ and ${\rm d}\omega/{\rm d}I < 0$ is chosen arbitrarily.
		Regularization constant was increased until power penalty reached -25~dB.
		(c)~3-step prediction with TPA+FCD case at an optimal input power, here NRMSE is 0.15 and $L_{\rm p} \approx -22$~dB.
		(d)~Trajectories of linearized Jacobian $J_{km}$ for (c).
	}
	\label{fig:mg-bundle}
\end{figure}

%
Nevertheless, the reservoir used is time-continuous, we then convert $\xi(t_n)$ into a continuous signal by a zero-hold interpolation.
This signal is used to modulate an amplitude of optical carrier that is injected into reservoir as an input.
To find the output $z_k(t)$ are sampled at $t_n$.
Due to the nature of $\xi(t_n)$, the Fourier spectrum of optical input is uneven~(see Figure~\ref{fig:mg-fourier}).
As a result, the excitation of supermodes is not uniform with a non-trivial impact on the reservoir.
To visualize the impact of nonlinearities, we consider a linearized Jacobian of the reservoir
\begin{equation}
	J_{km} = \rm{diag}\left({\rm i}\bomega - \frac{\bGamma^{\rm o} + \bGamma^{\rm e}}{2}\right)_{km}
	+ \hat M^{\mu}_{km}
	+ \hat M^{\kappa}_{km}
	+ \left [
		- \frac{1}{2}\frac{\beta_2 c^2}{n^2 V^{{\rm TPA}}}I_k
		+ {\rm i}\frac{{\rm d}\omega}{{\rm d}N}N_k
		+ {\rm i}\frac{{\rm d}\omega}{{\rm d}I}I_k
	\right]\delta_{km},
	\label{eq:mg-eig-traj}
\end{equation}
where last three terms correspond to TPA, FCD and the Kerr effect, with $\delta_{km}$ as the Dirac delta.
In Figure~\ref{fig:mg-bundle}(d) we show trajectories of $\hat J$ eigenvalues of the system while driven by the input.
As expected, TPA increases eigenloss and modifies eigenfrequencies, most likely due to modifications to the spatial shape of the supermodes.
With FCD these modifications become notably stronger, and in the case of Kerr effect they are the dominant feature.
This behaviour is notably different from ESNs~\cite{ozturk2007}.
The inequality of nonlinearity strength is also evident.
A supermode near a Fourier peak with a strong connection to an input waveguide is subject to several times stronger nonlinearity than others.
This poses a challenge for FCD and the Kerr effect as a nonlinearity concentration in a few supermodes leads the reservoir to chaos before nonlinearity in other supermodes becomes strong enough.

Besides the obvious relevance of the NRMSE for performance, the output signal power also becomes a concern in noisy systems.
Even though a noise was not present in simulations, we also consider a power penalty as an additional performance metric:
$
	L_{\rm p} = 10 \log(\langle |y(t)|^2 \rangle_t / \langle |s(t)|^2 \rangle_t),
$
where $\langle \dots \rangle_t$ is an average over time.

In Figure~\ref{fig:mg-bundle}(a) a considerable improvement with all three nonlinearities is seen.
However, with frequency-shifting nonlinearities the performance is lost after the normalized average input power approaches unity.
This is consistent with Section~\ref{sec:dimensionality-nonlinear} as with a loss of consistency the training becomes ineffective.
However, in Figure~\ref{fig:mg-bundle}(b) we see that before the consistency is lost, the output power is higher compared to TPA, which is likely due to an increase of dimensionality (see Figure~\ref{fig:dim-cons}).
For a $8\times 3$ reservoir with TPA and FCD a NRMSE of $1.5\cdot 10^{-1}$ was achieved (see Figure~\ref{fig:mg-bundle}(c)), while for TPA $8\cdot 10^{-2}$, considering power penalty restriction.
For comparison, a performance baseline is ${\rm NRMSE}(u_n, u_{n+3}) \approx 1.1$, an all-optical time-multiplexed reservoir with 330 virtual neurons based on a semiconductor laser experimentally achived NRMSE of 0.23~\cite{bueno2017}, a simulated time-multiplexed reservoir based on a nonlinear MRR with 25 virtual nodes had NRMSE of $7.2\cdot10^{-2}$~\cite{donati2022}.
However, an ESN with 400 neurons outperforms by a large margin, predicting approximately 20 steps ahead with a NRMSE of $1.2\cdot10^{-4}$~\cite{jaeger2001}.
Nevertheless, the proposed reservoir performs such computation fully-optically in real time.

\begin{figure}[hbtp!]
	\centering
	\includegraphics[width=\textwidth]{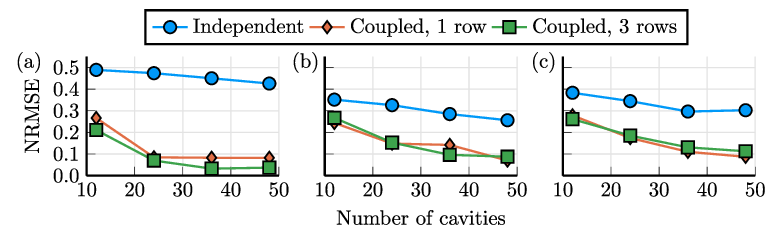}
	\caption{
		Scalability of 3-step prediction performance with various geometries and (a) TPA (b) TPA with FCD and (c) the Kerr effect.
		For each point the regularization constant was increased until power penalty reached -25~dB.
		Minimum NRMSE (a) 0.037 (b) 0.070 (c) 0.087.
	}
	\label{fig:mg-scaling}
\end{figure}

Simulations have shown that prediction performance improves with the number of cavities (see Figure~\ref{fig:mg-scaling}) for $N_\perp = 1$ or 3 and for all three nonlinearities.
With FCD and the Kerr effect $N_\perp = 1$ and $N_\perp = 3$ reservoirs have shown similar performance, while for TPA $N_\perp = 3$ was better than $N_\perp = 1$.
The lowest error with an NRMSE of 0.037 and $L_p$ of -25~dB is achieved for 48 cavities with $N_\perp = 3$.
We also note the importance of the direct coupling by comparing our system with a set of independent cavities, a linear regime of which was considered by~\cite{hermans2010}.
For a fair comparison, resonances of cavities were set to resonances of supermodes of a corresponding $N \times 1$ reservoir.
With all nonlinearities coupled cavities provided considerably better performance, likely due to a nonlinearity-induced supermode mixing.

\subsection{Optical signal recovery}
\label{sec:nlcheq}
\begin{figure}[hbtp!]
	\centering
	\includegraphics[width=\textwidth]{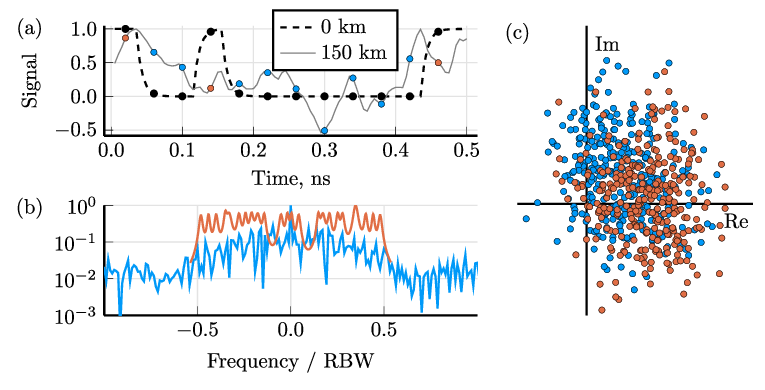}
	\caption{
		(a) Timetraces of real part of original signal (dashed line) and distorted signal after 150~km of optical fiber (solid line).
		Black dots show bit middle and act as training target, blue and orange are only for visualization and show distorted signal samples with color corresponding to each symbol.
		(b) Normalized Fourier spectrum of distorted signal after 150~km of fiber (blue) with $8\times 3$ reservoir resonances overlaid (orange).
		(c) Constellation of distorted signal samples after 150~km of fiber, colors correspond to (a).
	}
	\label{fig:nlcheq-bundle}
\end{figure}

An optical signal propagating through an optical fiber is subject to the chromatic dispersion and the Kerr nonlinearity, among other effects.
Electric field $E(z,t)$ in the fiber can be described with a nonlinear Schrödinger equation~\cite{agrawal2007}:
\begin{equation}
	{\rm i} \frac{\partial E}{\partial z} + {\rm i} \frac{\alpha_{{\rm loss}}}{2}E - \frac{\beta_2}{2}\frac{\partial^2 E}{\partial t^2} + \gamma |E|^2E= 0,
	\label{eq:nlcheq}
\end{equation}
where we choose $\alpha_{\rm loss} = 0.2$~dB/km, $\beta_2 = 21.7$~ps$^2$/km, $\gamma=1.3$~W$^{-1}$/km, $E(0,t) = s_0(t)$ and $s_0(t)$ is the transmitted signal.
Because of distortion, after propagating through a fiber of length $L$ the signal at the receiver side $E(L,t) = s(t)$ differs from $s_0(t)$. 
Examples of the distortion are shown in Figure~\ref{fig:nlcheq-bundle}(a,b).
If $s_0(t)$ is a carrier of encoded symbols, the reservoir's task is to recover them from $s(t)$.

We generate a random sequence $s_{{\rm 0}}(t_{n})$ of symbols encoded with on-off keying at 25~GHz with 8 samples per bit and filter with a first order low-pass filter with $0.6\cdot25$~GHz cutoff frequency (see Figure~\ref{fig:nlcheq-bundle}(a)).
The launch power is 10~dBm.
Then Eq.~\ref{eq:nlcheq} is solved to obtain distorted signal samples $s(t_{n})$, which are linearly interpolated to $s(t)$ and injected into the reservoir.
The reservoir parameters are the same as in the Mackey-Glass task (see Table~\ref{tab:mg}) except for $\mu = 40$~GHz, chosen according to the $s(t)$ spectrum (see Figure~\ref{fig:nlcheq-bundle}(b)) and nonlinear effects disabled.
As in the previous section, $\mu$ is increased for $N_\parallel \times 1$ reservoirs to equalize reservoir bandwidth.
For RC training we set $y^{\rm tgt}(t) = s_0(t - \tau)$ where $\tau$ is an output delay and sample $s_0(t)$, $s(t)$ and $z_k(t)$ once per bit in its middle (see dots in Figure~\ref{fig:nlcheq-bundle}(a)).
For RC testing $y(t)$ and $y^{\rm {tgt}}(t)$ are similarly sampled to produce $y_n$ and $y_n^{\rm tgt}$ and a linear classifier $C$ is used.
Performance is measured with the Symbol Error Rate (SER): ${\rm SER}(y, y^{\rm tgt}) = \sum_n\mathds{1}\left[C(y_n) \ne C(y_n^{\rm tgt})\right] / N_{\rm t}$, where $N_{\rm t}$ is a number of transmitted symbols.
However, SER cannot be lower than $1 / \min_{c}\left\{\sum_n\mathds{1}[C(y_n^{\rm tgt}) = c]\right\}$, which in our case is approximately $2/N_{\rm t}$.

\begin{figure}[hbtp!]
	\begin{center}
		\includegraphics[width=\textwidth]{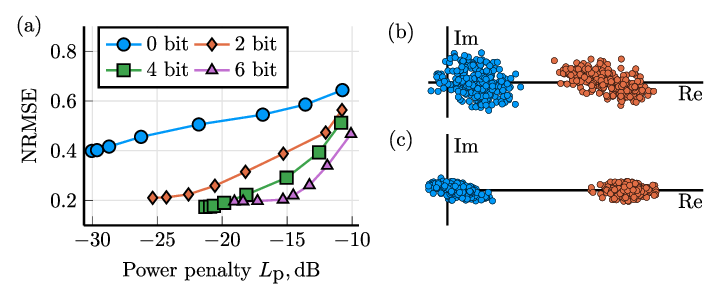}
	\end{center}
	\caption{
		(a) Effect of output delay (shown on legend) on output symbol separability measured with NRMSE at a given power penalty, controlled with regularization parameter.
		Here $L = 150$~km and signal is processed with a $8 \times 3$ linear reservoir.
		(b)(c) Constellations of output symbols at $L_{\rm p} \approx -20$~dB with 0 and 4 bit delay respectively.
		Figure~\ref{fig:nlcheq-bundle}c shows constellation of input signal samples.
	}
	\label{fig:nlcheq-delay}
\end{figure}

The presence of an output delay is motivated by group velocity dispersion that causes different input signal frequencies to arrive to a receiver at different times.
By delaying $y^{\rm tgt}(t)$ we allow the reservoir to accumulate more information about the current symbol before producing an output.
Simulations show that it leads to a better separation of output samples at a given power penalty.
For used fiber parameters a delay of 4 bits is adequate and is used throughout the section (see Figure~\ref{fig:nlcheq-delay}).

\begin{figure}[hbtp!]
	\begin{center}
		\includegraphics[width=\textwidth]{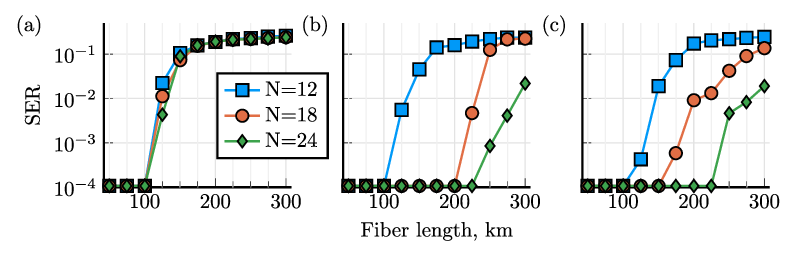}
	\end{center}
	\caption{
		Performance scaling of optical signal recovery w.r.t. reservoir size.
		The reservoir is (a) $N$ independent MRRs (b)~$N_\parallel \times 1$ grid (c)~$N_\parallel \times 3$ grid of linear MRRs.
		For each computation the regularization parameter is increased until the power penalty reached -25~dB.
	}
	\label{fig:nlcheq-scaling}
\end{figure}

A larger reservoir is capable of handling a distortion from a longer fiber (see Figure~\ref{fig:nlcheq-scaling}).
The performance of a $N_\parallel \times 1$ grid of cavities was comparable to a $N_\parallel \times 3$ grid and outperformed independent cavities.

\section{Discussion}

\begin{figure}[hbtp!]
	\centering
	\includegraphics[width=0.3\textwidth]{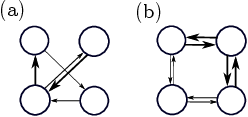}
	\caption{
		Neuron connectivity graph of (a) a typical ESN and (b) a photonic RC with a grid of cavities.
		Connection strength is represented with arrow thickness.
	}
	\label{fig:conn-scheme}
\end{figure}

The proposed setup can successfully be used as a RC.
However, there are notable differences that need discussion.
First of all, an ESN connection matrix is typically random and sparse~\cite{lukovsevivcius2012}, and two neurons are unlikely to have a connection equally strong in both directions.
In the photonic case if two cavities are directly coupled, the connection is reciprocal and coupling coefficients are negative complex conjugate of each other~(see Figure~\ref{fig:conn-scheme}).

Moreover in ESNs a nonlinear function encompasses the input and the neuron connection terms~(see Eq.~\ref{eq:reservoir-eq}).
In our reservoir, however, a nonlinearity depends only on the current reservoir state~(see Eq.~\ref{eq:cavity-eq}).
Nonlinearities themselves are differrent as well.
Typically, in ESNs a sigmoid or hyperbolic tangent are used.
They add a saturation of neuron activation, which is, effectively, an increase of loss.
During a typical computation, ESN's linearized Jacobian eigenvalues largely move in the direction of the z-plane origin~(corresponding to an increase of loss) and go back with, perhaps, slight frequency deviations~\cite{ozturk2007}.
In the photonic case with TPA as the only nonlinearity, the behaviour is similar, but with FCD and the Kerr effect the trajectories of eigenvalues change considerably~(see Figure~\ref{fig:mg-bundle}(d)).
Nonetheless, all three nonlinearities positively contribute to performance~(see Figure~\ref{fig:mg-bundle}(a)).


In Section~\ref{sec:dimensionality-linear} a relation between cavity parameters and their effect on reservoir dimensionality was proposed.
Whether these requirements could be met with the available technology needs to be discussed.
First, consider that $\Gamma^{{\rm o}} < \Gamma^{{\rm S}}$, $F_\Gamma \approx 1$ and ${\rm RBW} \le {\rm BW}$
\begin{equation}
	N \frac{\Gamma^{{\rm o}}}{2\pi} \le N \frac{\Gamma^{{\rm S}}}{2\pi} \approx N F_\Gamma \frac{\Gamma^{{\rm S}}}{2\pi} = {\rm RBW} \le {\rm BW} \Rightarrow \Gamma^{{\rm o}} \lessapprox \frac{2\pi}{N}{\rm BW}.
	\label{eq:gammao-fsr}
\end{equation}
Another requirement demands of the reservoir to remember previous inputs: an optical timescale should be larger than an input signal timescale, i.e.
\begin{equation}
	\frac{1}{{\rm BW}} < \frac{2}{\Gamma^{{\rm S}}} < \frac{2}{\Gamma^{{\rm o}}} \Rightarrow \Gamma^{{\rm o}} < 2{\rm BW},
\end{equation}
which is already satisfied by Eq.~\ref{eq:gammao-fsr}.
For example, consider a reservoir with 20 cavities.
To process a signal with ${\rm BW} = 20$~GHz, an approximate lower bound for $\Gamma^{{\rm o}}$ would be $2\pi$~GHz, which would correspond to an intrinsic Q-factor of $\omega_{{\rm 0}} / \Gamma^{{\rm o}} \approx 0.2 \cdot 10^6$.
This is rather high, but not impossible.
An intrinsic Q-factor of $0.7\cdot 10^6$ was demonstrated in passive GaAs PhCs with $10^6$ considered achievable~\cite{combrie2008}.
In an AlGaAs MRR a Q-factor of $1.5\cdot 10^6$ was demonstrated~\cite{chang2020} and in a silicon PhC a Q-factor of more than $11\cdot 10^6$~\cite{asano2017}.
The low-loss silicon nitride allowed for even higher Q-factor of $37\cdot 10^6$ in a MRR~\cite{ji2017}.

Next, consider a nonlinearity timescale.
TPA and the Kerr effect respond almost instantaneously to a change of electric field, similar to an ESN.
However, FCD is also affected by the free electron recombination rate (see Figure~\ref{fig:refpower}(a)), which should be comparable to a rate at which of optical signal in a cavity changes, i.e. half of its bandwidth.
In our case each cavity belongs to multiple supermodes and thus its bandwidth approaches the ${\rm RBW}$, which is close to ${\rm BW}$.
The electron recombination rate of silicon MRRs was shown to be approximately 2 GHz~\cite{preble2005}, i.e. ${\rm BW} \lessapprox 4$~GHz would be supported.
The recombination can be accelerated to almost 20~GHz with a reverse-biased p-i-n junction embedded in a cavity~\cite{preble2005}.
Due to a higher surface area, in PhCs the recombination can reach 10~GHz for silicon~\cite{leonard2002} and 100~GHz for GaAs~\cite{combrie2008}.

\begin{figure}
	\begin{center}
		\includegraphics[width=\textwidth]{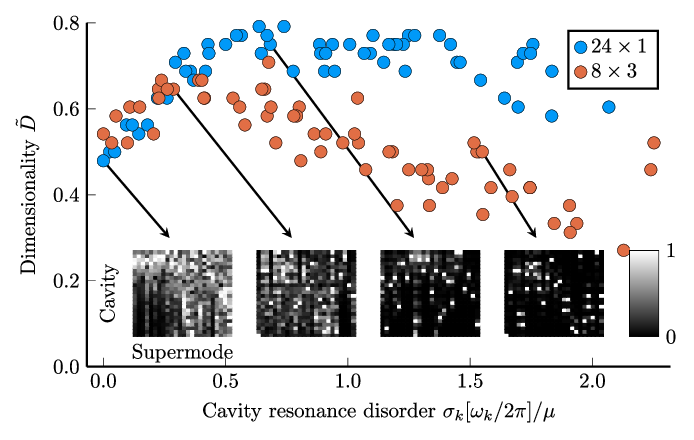}
	\end{center}
	\caption{
		Impact of cavity resonance frequency disorder on dimensionality.
		Reservoir parameters given in Table~\ref{tab:mg}, ${\rm BW} > {\rm RBW}$.
		Each dot corresponds to a $N_\parallel \times N_\perp$ (shown on legend) linear MRR reservoir with resonances $\omega_k \in \mathcal{N}(\omega_0, \sigma_\omega)$ with various $\sigma_\omega$.
		Insets represent supermode intensity distribution over cavities $\hat \Lambda$.
	}
	\label{fig:resonance-disorder}
\end{figure}

Until this point we assumed that cavities's resonances are aligned $\forall k=1..N~\omega_k = \omega_0$.
However, this might not be the case due to fabrication tolerances.
In order for directly coupled cavities to interact, resonances should be sufficiently close.
A weak intercavity interaction leads to two issues, the first being a dimensionality reduction when some cavities are not coupled to the input waveguide directly (e.g. for $N_\perp > 1$).
Since such cavities can only receive signals through other cavities, without an interaction they stop receiving signals and hence cannot contribute to dimensionality.
This is demonstrated in Figure~\ref{fig:resonance-disorder} for $8 \times 3$ and $24 \times 1$ linear MRR reservoirs.
As a resonance disorder is increased, at first $\tilde D$ increases as the supermode overlap is alleviated, but decreases afterwards.
The $24 \times 1$ reservoir is affected less than the $8 \times 3$ one, since in the former 12 cavities are coupled to the waveguide (due to MRR mode direction matching, see Figure~\ref{fig:dims}(b)), but only 4 in the latter.
If MRR mode direction was not considered and both MRR modes were excited, all cavities of a $24 \times 1$ MRR reservoir are coupled to a waveguide.
In that case a further increase of resonance disorder simply increases $F_\Gamma$ and $\tilde D$ will eventually reach unity.
A similar outcome is expected for a PhC reservoir.

The second issue is a weakening of supermode mixing, i.e. delocalization.
When cavities interact well, supermodes span multiple cavities, otherwise, they isolate in individual cavities.
Because of that, a nonlinearity of an individual cavity can only affect a single supermode.
As was shown in Figure~\ref{fig:mg-bundle}(d), in this case nonlinearity could not contribute to computing performance as much.
Here, $N_\perp > 1$ cases could be affected to a lesser degree than $N_\perp = 1$, since for $N_\perp > 1$ each cavity has 2-4 neighbouring cavities, while only 1-2 for $N_\perp = 1$ and the probability of resonance mismatch with all neighbours is lower with more neighbours.
The weakening of supermode mixing is shown in inset of Figure~\ref{fig:resonance-disorder}.
Each heatmap represents a matrix $\hat \Lambda$ where $\Lambda_{km}$ is an intensity of $m$-th supermode in the $k$-th cavity.
For clarity, each $m$-th column of $\hat \Lambda$ is normalized to unity.
We see that when $\sigma_k[\omega_k/2\pi] < 0.5\mu$, supermodes span multiple cavities, but only a few otherwise.
The resonance disorder can be mitigated with the thermo-optic effect, but for more cavities it becomes challenging.
In addition to that, a close proximity of cavities results in a strong thermal crosstalk.
However, the resonance disorder can be less severe when cavities are spatially close~\cite{dodane2018}.



\section{Conclusion}
We have applied a network of directly coupled nonlinear nanophotonic cavities for a reservoir computing.
Such coupling allows for a lower chip footprint by several orders of magnitude, excluding readout.
We have derived general conditions under which such a system attains a high dimensionality, an important property for RC.
The role of nonlinear effects present in integrated microcavities, namely TPA, FCD and the Kerr effect were studied.
While their impact on the reservoir dimensionality and the consistency is different, all three nonlinearities positively contributed to the performance in Mackey-Glass prediction task.
The reservoir was also successful in recovery of OOK-encoded optical signal.
In both cases the reservoir has shown a fully optical autonomous real-time computation in the GHz range.


Several potential improvements of this design could be proposed.
For example, using multiple cavity resonances to operate on multiple channels at the same time.
An another option is to overcome dimensionality limitation due to a finite Q-factor.
If a reservoir has an upper limit of $N$ cavities, one could consider using two such reservoirs with one receiving a delayed copy of input.
This way the dimensionality of the setup could be expected to increase two-fold.



\section*{Acknowledgments}
This work was supported by European Union through the Marie Skłodowska-Curie Innovative Training Networks action, POST-DIGITAL project number 860360.




\appendix

\section{Derivation of waveguide-related differential equation matrices}
\label{sec:wg-cav-matrix}
\begin{figure}[htbp!]
	\begin{center}
		\includegraphics[width=0.4\textwidth]{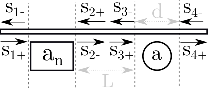}
	\end{center}
	\caption{Scheme of waveguide-cavity interaction. A rectangle with $a_n$ represents a row of $n$ cavities coupled to the waveguide. }
	\label{fig:waveguide-coupling}
\end{figure}

A mode amplitude reduces exponentially outside a resonator, therefore we can assume that each cavity-waveguide interaction happens in a restricted region between two reference planes separated by $d_k$ (dashed lines in Figure~\ref{fig:waveguide-coupling})~\cite{manolatou1999}.
For a single cavity coupled to a waveguide the equations are known~\cite{manolatou1999}:
\begin{equation}
	\frac{{\rm d}a}{{\rm d} t} = -\left(\frac{|\kappa_{1}|^2}{2} + \frac{|\kappa_{2}|^2}{2} \right) a + \kappa_{1}s_{1+} + \kappa_{2}s_{2+},
\end{equation}
\begin{equation}
	\cases{
		s_{1-} = e^{-i\beta d} (s_{2+} - \kappa_{2}^*a) \\
		s_{2-} = e^{-i\beta d} (s_{1+} - \kappa_{1}^*a)
	}.
\end{equation}
Omitting waveguide-induced loss and generalizing for a row of $n$ cavities coupled to one waveguide
\begin{equation}
	\frac{{\rm d}\vec{a_n}}{{\rm d} t} = \hat M^n \vec{a_n} + \vec{R_1^n}s_{1+} + \vec{R_2^n}s_{2+},
\end{equation}
\begin{equation}
	\epmatrix{s_{1-}\cr\\s_{2-}} =
\left(\ematrix{0 & T_1^n \cr\\ T_2^n & 0 }\right)\epmatrix{s_{1+}\cr\\s_{2+}} +
\left(\ematrix{\vec{Q_1^n} \cr\\ \vec{Q_2^n} }\right)\vec{a_n},
\end{equation}
where $\hat M^n$, $\vec{R_1^n}$ and $\vec{R_2^n}$ are assumed to be known.
Adding a new cavity to the row
\begin{equation}
	\frac{{\rm d}a}{{\rm d} t} = \kappa_{1}s_{3+} + \kappa_{2}s_{4+},
\end{equation}
\begin{equation}
	\cases{
		s_{3-} = e^{-i\beta d} (s_{4+} - \kappa_{2}^*a) \\
		s_{4-} = e^{-i\beta d} (s_{3+} - \kappa_{1}^*a) \\
		s_{3+} = e^{-i\beta L} s_{2-} \\
		s_{2+} = e^{-i\beta L} s_{3-}
	},
\end{equation}
The goal is to find the matrix representation of a system with $n+1$ cavities based on a system with $n$ cavities.
We first eliminate the flux between the cavities
\begin{equation}
	\cases{
	s_{2+} = e^{-i\beta L}e^{-i\beta d} (s_{4+} - \kappa_{2}^*a) \\
	s_{3+} = e^{-i\beta L} \left(T_2^n s_{1+} + \vec{Q_2^n}\vec{a_n} \right)
}.
\end{equation}
Then, cavity dynamics can be rewritten using only new inputs $s_{1+}$ and $s_{4+}$:
\begin{equation}
	\cases{
	\frac{{\rm d}\vec{a_n}}{{\rm d} t} = \hat M^n \vec{a_n} + \vec{R_1^n}s_{1+}
	+ e^{-i\beta L}e^{-i\beta d}\vec{R_2^n}s_{4+}
	-e^{-i\beta L}e^{-i\beta d}\vec{R_2^n}\kappa_{2}^*a, \\
	\frac{{\rm d}a}{{\rm d} t} = e^{-i\beta L} T_2^n \kappa_{1}s_{1+} + e^{-i\beta L} \vec{Q_2^n}\kappa_{1}\vec{a_n} + \kappa_{2}s_{4+},
}
\end{equation}
Consequently, new outputs are:
\begin{equation}
	\eqalign{
		s_{1-}
	&= T_1^n s_{2+} + \vec{Q_1^n}\vec{a_n} = \\
	&= T_1^n e^{-i\beta L} \left[ e^{-i\beta d} (s_{4+} - \kappa_{2}^*a) \right] + \vec{Q_1^n}\vec{a_n} = \\
	&= T_1^n e^{-i\beta L}e^{-i\beta d} s_{4+} + \vec{Q_1^n}\vec{a_n} - T_1^n e^{-i\beta L} e^{-i\beta d} \kappa_{2}^* a,
}
\end{equation}
\begin{equation}
	\eqalign{
		s_{4-}
		&= e^{-i\beta d} (e^{-i\beta L} s_{2-} - \kappa_{1}^*a) = \cr
		&= e^{-i\beta d} \left[ e^{-i\beta L} \left(T_2^n s_{1+} + \vec{Q_2^n}\vec{a_n}\right) - \kappa_{1}^*a \right] = \cr
		&= e^{-i\beta d} e^{-i\beta L} T_2^n s_{1+} + e^{-i\beta d} e^{-i\beta L} \vec{Q_2^n}\vec{a_n} - e^{-i\beta d} \kappa_{1}^*a.
	}
\end{equation}
Therefore, the differential equation and the output matrices can be build incrementally:
\begin{equation}
	\eqalign{
		\frac{\rm d}{{\rm d}t}\epmatrix{\vec{a_n}\\a}
	&= \left(\ematrix{ \hat M^n & -e^{-i\beta L}e^{-i\beta d}\vec{R_2^n}\kappa_{2}^* \cr\\ e^{-i\beta L}\vec{Q_2^n}\kappa_{1} & 0 }\right)\epmatrix{\vec{a_n}\\a} + \cr\\
	&+ \left(\ematrix{ \vec{R_1^n} & e^{-i\beta L}e^{-i\beta d}\vec{R_2^n} \cr\\ e^{-i\beta L} T_2^n \kappa_{1} & \kappa_{2} }\right)\epmatrix{s_{1+} \cr\\ s_{4+}},
}
\end{equation}
\begin{equation}
\eqalign{
	\epmatrix{s_{1-}\cr\\s_{4-}}
    &= \left(\ematrix{0 & e^{-i\beta d} e^{-i\beta L}T_1^n \cr\\ e^{-i\beta d} e^{-i\beta L} T_2^n & 0 }\right)\epmatrix{s_{1+}\cr\\s_{4+}} + \cr\\
    &+ \left(\ematrix{ \vec{Q_1^n} & - e^{-i\beta L} e^{-i\beta d}T_1^n \kappa_{2}^* \cr\\ e^{-i\beta d} e^{-i\beta L} \vec{Q_2^n} & - e^{-i\beta d} \kappa_{1}^* }\right)\epmatrix{\vec{a_n}\cr\\a}.
}
\end{equation}


\section{Comparing evanescent and dissipative coupling}

Consider a linear case of two directly coupled lossless cavities.
Their state will be described with Eq.~\cite{haus1984}:
\begin{equation}
	\frac{{\rm d}}{{\rm d}t}\left(\ematrix{a_1 \cr a_2 \cr}\right) = \left(\ematrix{0 & \mu \cr -\mu^* & 0}\right)\left(\ematrix{ a_1 \cr a_2 \cr}\right),
	\label{eq:direct-coupling}
\end{equation}
where $\mu$ is a coupling coefficient and $^*$ is a complex conjugation.
Then $a_{1,2} \propto \exp(\pm i|\mu| t)$, i.e. no loss was introduced regardless of $\mu$.
Comparing to a similar case of waveguide coupling of two PhCs in simplified form (see \ref{sec:wg-cav-matrix}):
\begin{equation}
	\frac{{\rm d}}{{\rm d}t}\left(\ematrix{a_1 \cr a_2 \cr}\right) =
	\left(\ematrix{-|\kappa|^2 & |\kappa|^2 \exp({\rm i}\varphi_{12}) \cr |\kappa|^2 \exp({\rm i}\varphi_{21}) & -|\kappa|^2 \cr}\right)\left(\ematrix{a_1 \cr a_2}\right),
	\label{eq:}
\end{equation}
depending on $\varphi_{12}$ and $\varphi_{21}$ there could be an interaction ranging from the one similar to direct coupling to its total absence, i.e. light escapes from cavities to the input waveguide.
In practice, however, phases are effectively random.
As a result, we choose the direct coupling as a neuron connectivity method.

\section{Estimation of reference power for a single cavity}
Assuming a monochromatic excitation with an input power $P_{{\rm NL}}$ and a steady-state, a power that the cavity receives should be equal to a power lost:
\begin{equation}
	P_{{\rm NL}} = -\frac{d|a|^2}{{\rm d}t} = -\frac{da}{{\rm d}t}a^* - {\rm c.c.} = \left[\Gamma^{{\rm S}} + \frac{\beta_2 c^2}{n^2 V^{{\rm TPA}}}|a|^2\right]|a|^2.
\end{equation}
For a TPA-dominated case the reference power is found when $\Gamma^{{\rm TPA}}(a) = \Gamma^{{\rm S}}$, then
\begin{equation}
	|a|^2 = \frac{\Gamma^{{\rm S}}n^2 V^{{\rm TPA}}}{\beta_2 c^2}.
\end{equation}
For a FCD-dominated case in a steady state (assuming that a detuning does not impact the power absorbed by the cavity):
\begin{equation}
	\Delta\omega = \frac{\partial \omega}{\partial N} N = \frac{\partial \omega}{\partial N} \frac{\beta_2c^2}{n^2V^{{\rm TPA}}}\frac{1}{2\hbar\omega_{{\rm 0}} V^{\rm{c}}\Gamma^{{\rm c}}}|a|^4 = \Gamma^{{\rm S}},
\end{equation}
\begin{equation}
	|a|^2 = \left(\frac{\Gamma^{{\rm S}}\Gamma^{{\rm c}}n^2V^{{\rm TPA}}2\hbar\omega_{{\rm 0}} V^{\rm{c}}\Gamma^{{\rm c}}}{(\partial\omega/\partial N)\beta_2c^2}\right)^{1/2}.
\end{equation}

\printbibliography

\end{document}